# Determining the grain orientations of battery materials from electron diffraction patterns using convolutional neural networks

Jonas Scheunert,[1,2] Shamail Ahmed,[1,2] Thomas Demuth,[1,2] Andreas Beyer,[1,2]

Sebastian Wissel[3], Bai-Xiang Xu[3] and Kerstin Volz[1,2*]

[1] mar.quest | Marburg Center for Quantum Materials and Sustainable Technologies, Philipps-Universität Marburg, Hans-Meerwein Straße 6, 35032 Marburg, Germany

[2] Department of Physics, Philipps-Universität Marburg, Hans-Meerwein Straße 6, 35032 Marburg, Germany

[3] Mechanics of Functional Materials, Institute of Materials Science, TU Darmstadt, Otto-Berndt Straße 3, 64287 Darmstadt, Germany

*Corresponding author: kerstin.volz@physik.uni-marburg.de

## Abstract

Polycrystalline materials have numerous applications due to their unique properties, which are often determined by the grain boundaries. Hence, quantitative characterization of grain as well as interface orientation is essential to optimize these materials, particularly energy materials. Using scanning transmission electron microscopy, matter can be analysed in an extremely fine grid of scan points via electron diffraction patterns at each scan point. By matching the diffraction patterns to a simulated database, the crystal orientation of the material as well as the orientation of the grain boundaries at each scan point can be determined. This pattern matching approach is highly time intensive. Artificial intelligence promises to be a very powerful tool for pattern recognition. In this work, we train convolutional neural networks (CNNs) on dynamically simulated diffraction patterns of $LiNiO_2$, an important cathode-active material for Lithium-ion batteries, to predict the orientation of grains in terms of three Euler angles for the complete fundamental orientation region. Results demonstrate that these networks outperform the conventional pattern matching algorithm with increased accuracy and efficiency. The increased accuracy of the CNN models can be attributed to the fact that these models are trained by data incorporating dynamical effects. This work is the first attempt to apply deep learning for analysis of electron diffraction data and enlightens the great potential of ML to accelerate the analysis of electron microscopy data, toward high-throughput characterization technique.

## Introduction

Many modern functional devices like batteries, solar cells or fuel cells consist of an agglomeration of individual crystallites. The efficiency of the respective device depends largely on the orientation of the grains and especially the boundaries or interfaces between them. Accordingly, characterization methods to derive the grain orientations on a μm to nm level are needed. To this end, the commonly applied technique is analysing electron backscatter diffraction (EBSD) patterns based on the Laue method (1) in a scanning electron microscope (SEM). Even higher spatial resolution can be achieved by transmission electron microscopy (TEM). Here, the orientation of a crystal can be determined by indexing the recorded spot-like diffraction patterns. The diffraction patterns are acquired at each scan point resulting in a four-dimensional data set (4D-STEM). The experimental diffraction patterns are then compared to a large reference library of simulated diffraction patterns via a cross-correlation algorithm (2) and a two-dimensional map of the crystallographic orientation of the material is created. (3) Typically, the simulated diffraction patterns in this method are calculated by applying the kinematic approximation to speed up the simulation process. However, this neglects dynamical effects like the intensity modulation of diffraction spots with respect to sample thickness and excitation error. Accordingly, the accuracy of such an orientation mapping can be significantly enhanced by applying precession electron diffraction (PED) (4) to minimize the impact of dynamic effects in the experimental data. Although somewhat reliable, the orientation mapping is less robust on inhomogeneous samples, when, e.g., the thickness of the sample or the background in the diffraction pattern changes across the specimen. All in all, these methods are computationally expensive and therefore still are a time- and resource-intensive part of crystallographic analysis.

The emergence of artificial neural networks (ANNs) (5) provides a new method to analyse large amounts of data in a short amount of time. It also cuts down on the needed memory storage by effectively encoding the reference image library into the weights of the connections of the network architecture. Convolutional neural networks (CNNs) (6) have proven to be an effective method for analysing images.

Determining the crystal orientation by training CNNs on different data types of experimental data has been achieved by several groups before. Early achievements were reached by Jha et al. (7) by applying a convolutional neural network on EBSD patterns of polycrystalline nickel. Ding et al. (8) have recently developed a hybrid network consisting of a convolutional neural network combined with dictionary indexing and achieved high accuracy with a significant reduction in processing time for the EBSD patterns. For spatially resolved acoustic spectroscopy (SRAS) (9), a non-electron-based method, Patel et al. (10) have developed a fully connected network for indexing the orientation of nickel with data of this method. Experiments on using neural networks to index the orientation via spot-like transmission electron diffraction patterns have been carried out by Yuan et al. (11). They encoded the intensity of individual GaSb diffraction spots directly as values for nodes in a network. This method achieves very high accuracies, but only allows for small variations in the orientation to be analysed. To the best of our knowledge, there exists no study on the orientation mapping by utilizing TEM spot diffraction patterns covering the complete orientation range possible using CNNs.

In this work, we present our results on utilizing CNNs on TEM spot-like diffraction patterns to predict grain orientations covering the entire orientation range of a functional battery material. As example material, we chose $LiNiO_2$ because of its increasing relevance in battery material research. Our approach is to combine the three Euler angles, $\varphi 1$, $\varphi$, $\varphi 2$, of the Bunge notation (12) into a single label to be predicted by classification networks. By gradually breaking down the diffraction patterns into a low-dimensional information space through the architecture of a CNN, the networks also have the potential to identify decisive features more reliably than a simple pattern matching algorithm could do. For the training of the networks, we use dynamically simulated diffraction patterns using the Bloch wave algorithm, which further demarcates our approach from the template matching approach, where usually kinematic algorithms are employed to generate the reference database (3). Moreover, we investigate different strategies to distribute the training data in the orientation space.

## Methods

### Training data

In order to be able to predict crystal orientation accurately, a sufficient amount of training data for the neural network is needed. Therefore, we decided to use simulated diffraction patterns for the training, as they, unlike experimental data, can be created comparatively cheap and with very fine angular orientation steps. Most importantly, they provide perfect training labels as ground truth.

We simulate our data accounting for dynamical diffraction for various sample thicknesses at a fixed convergence semi-angle of 1.5mrad using the Bloch-wave formalism integrated in py4DSTEM (14). The acceleration voltage was set to 300kV and the maximum simulated scattering angle was set to 8mrad.

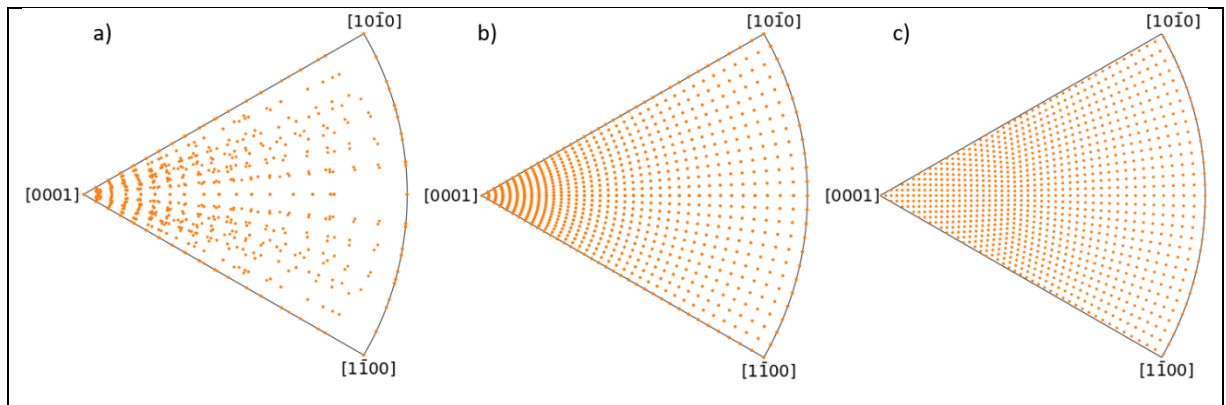

Fig. 1: Overview over the orientations used in the three training datasets. Variation of the first Euler angle is not shown in this representation. **a)** dataset with random orientations; **b)** dataset with individual steps of 2.5° for each Euler angle; **c)** dataset with equi-spaced orientations.

To investigate the influence of the distribution of the training data in the orientation space, we created three individual training datasets that differ in the way they cover the Euler angles, i.e. the different classes, in the orientation space (see Fig. 1). For all three datasets, the maximum values for the Euler angles are kept the same, reflecting the rhombohedral symmetry of $LiNiO_2$. Accordingly, $\varphi1$ varies from 0 to 360°, $\varphi$ from 0 to 90° and $\varphi2$ from 0 to 120°, respectively. As depicted in Fig. 1 (a), for the first dataset, the first angle ($\varphi1$) was varied in steps of 5°, whereas the second and third Euler angles ($\varphi$, $\varphi2$) were distributed randomly, which results in an inhomogeneous density of classes in the orientation space. For the second dataset, all three Euler angles were varied individually in steps of 2.5°. This results in a larger class density at low $\varphi1$ angles. To achieve a homogeneous density in the orientation space, $\varphi$ and $\varphi2$ angles of the third dataset were set in an equi-spaced manner with an angular resolution of 2° using the MTEX algorithm (15). The resulting homogeneous distribution of classes is visible in Fig. 1(c). In total, this results in about 70.000 classes for the first, 500.000 classes for the second and 150.000 classes for the third dataset, respectively. For the first dataset, we simulated diffraction patterns up to 100nm sample thickness in steps of 5nm. For the other two datasets, the step size was increased to 20nm in order to reduce the amount of data.

To increase the amount of training data further, augmentations were applied to the data. More importantly, by applying augmentations, the training results can be made more robust with respect to the different experimental conditions, like, e.g., the camera length or the camera model used for the acquisition. Therefore, to account for different camera sensitivities, the diffraction patterns were first rescaled to a range of 1 to a random value between 1e+3 and 1e+7 and then the logarithmically scaled. Furthermore, a Gaussian blurring filter was used to smoothen the diffraction patterns in order to account for a point spread function (PSF) of a camera. To further increase the robustness of the networks, variances in the investigated sample are augmented as well. To account for different extents of thermal diffuse scattering (TDS) and inelastic scattering, a Gaussian bell of intensity with random height and sigma were added. For 10% of the images diffuse rings of higher intensity around the central diffraction spot were added to mimic a varying amount of amorphous material that may be present on the sample surfaces from preparation. Last but not least, the diffraction patterns were randomly cropped and resized to 256x256 pixels. This accounts for different centering of the diffraction pattern on the camera and most importantly for different camera lengths used. For the template matching algorithms, the simulated database has to be calibrated beforehand by the user to match the experimental dataset as closely as possible. Neural networks provide the opportunity to eliminate the necessity of these calibrations by intentionally training them with diffraction patterns with varying camera lengths. From each simulated diffraction pattern, 30 different augmented images were created. As an example, Fig. 2 shows one simulated diffraction pattern (a) and two augmented training patterns (b) and (c) with different degrees of diffuse scattering and increasing camera lengths. For better visualisation, all images are scaled logarithmically. A fraction of 20% of the diffraction patterns generated are kept as a validation dataset, to check for overfitting during the training.

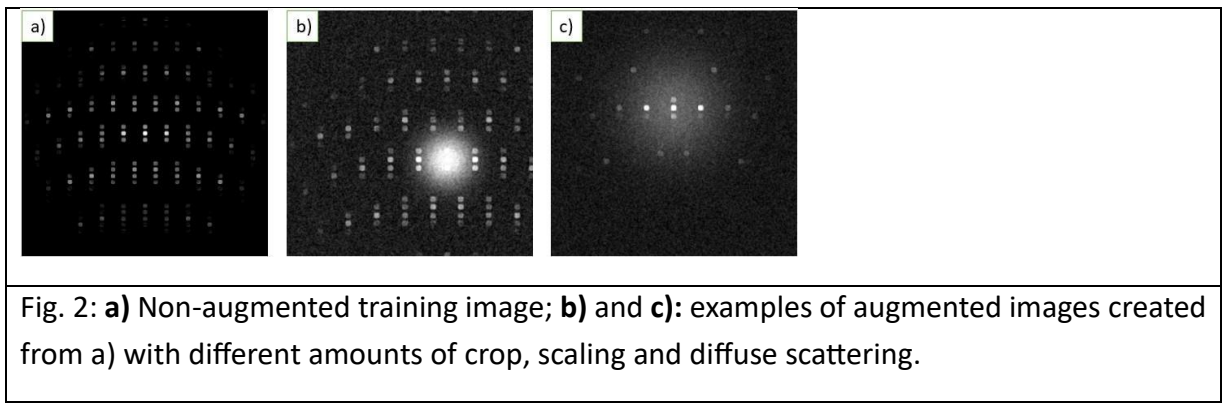

Fig. 2: **a)** Non-augmented training image; **b)** and **c):** examples of augmented images created from a) with different amounts of crop, scaling and diffuse scattering.

**Experimental data**

In order to prove the functionality of the networks, they were applied to evaluate an experimental dataset of $LiNiO_2$. For this a lamella of multiple single-crystal $LiNiO_2$ particles was prepared in a JEOL JIB 4601F focused ion beam microscope. To protect the lamella during the thinning process first a thin and then a thicker layer of tungsten were deposited on its surface with an electron-beam and a Ga-ion beam respectively. Using a micromanipulator needle the sample was then attached to a TEM grid and subsequently thinned using a Ga-ion beam. Finally, the lamella was polished with a 5 kV Ga-ion beam. The diffraction patterns were acquired in a JEOL JEM-3010 operating at 300kV, using a convergence semi-angle of 1.6mrad. The sample consists of multiple grains of different orientations. A NanoMegas P2010 system was used to precess and scan the electron beam across the sample. The precession angle was set to 0.6°. A total of 40.000 diffraction patterns were recorded in a grid of 200x200 scan points with a step size of 12.5nm using a TVIPS XF416(ES) camera. Fig. 3 shows a virtual dark-field overview image of the sample used (a) as well as an exemplary diffraction pattern (b). The diffraction pattern was logarithmically scaled for better visualisation. To create the labels for the experimental dataset, the ASTAR software was used to index the diffraction patterns (3) which is based on template matching of the experimental patterns to a set of kinematically simulated reference patterns.

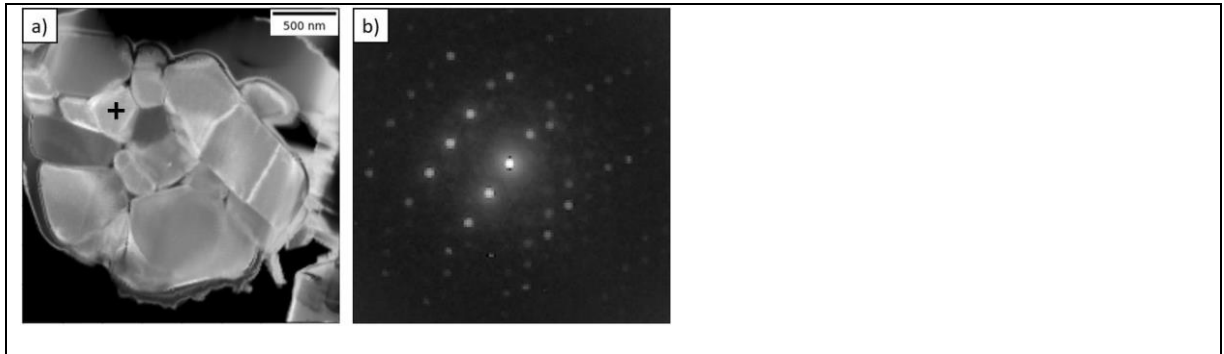

Fig. 3: **a)** darkfield overview image of the experimental dataset; **b)** diffraction pattern of the scan point marked in the black cross in the overview.

**Network architecture and training strategy**

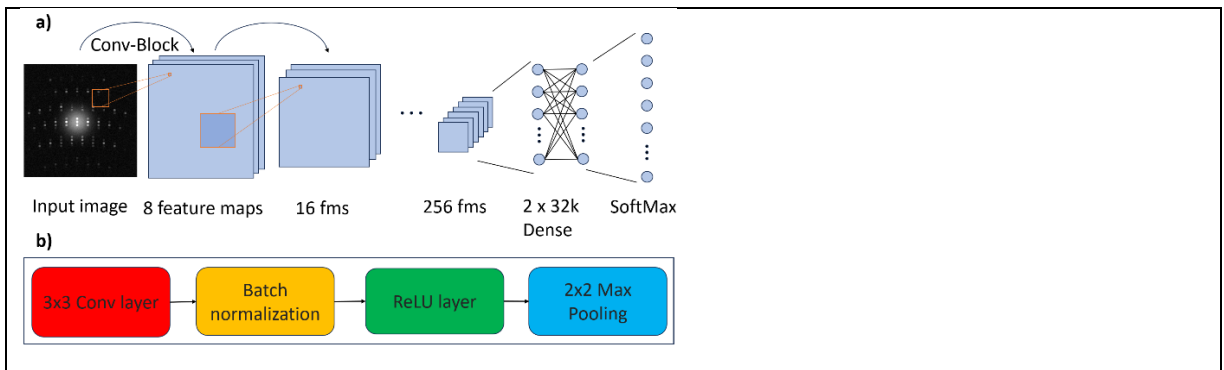

Fig. 4: **a)** General architecture of our convolutional neural networks; **b)** Setup of one convolutional block.

In this work, we employ classical convolutional neural networks because of their proven strength in image classification (16). We do not apply more complex architectures, like, e.g., the EfficientNet structure, since, due to our large number of classes, our networks already have a very high number of trainable parameters. Larger networks would slow down the training and evaluation process considerably. The general architecture of the networks used is shown in Fig. 4. The networks consist of multiple convolution blocks that gradually decrease the dimension of the diffraction patterns. Each block is made up of a convolution layer with a 3x3 filter kernel. This is followed by a layer for batch normalization (17) and one for ReLU activation (18). In the end, a 2x2 Max Pooling layer (19) is added. Using these blocks, we start with 8 feature maps and double the number with each convolution until we end up with 256 feature maps. After the convolution layers, two fully connected layers with an equal number of nodes are employed. For the first dataset, these layers have 6.400 nodes each. This number was reduced to 3.200 nodes for the other datasets, to reduce the number of trainable parameters. Dropout layers were included after each fully connected layer, in order to prevent overfitting. With these, 20% of the connections of those layers were deactivated during training. Lastly, another fully connected layer with as many nodes as classes in the dataset used was added. A Softmax layer was included as well to receive a predicted probability for each class.

All three networks were trained over 900 epochs, each consisting of 16.384 images. An adaptable learning rate was employed to maximize the achieved accuracy. If the validation loss is not decreasing after 20 epochs, the learning rate is reduced by 50%. The network trained on the first dataset achieved a training accuracy of 56%. The network trained on the second achieved an accuracy of 11.5%, while the third network reached 51% during training. It is obvious to see that the training accuracy decreases with an increase of the number of orientations. Given the highly similar looking diffraction patterns of orientations that differ only by a few degrees, it is also not surprising that the networks do not achieve higher accuracies. Fig. S1 in the supplementary material exemplarily shows the development of accuracy and loss of the second network during training.

## Results

In order to evaluate the networks, we compare their predictions of an experimental dataset with the orientations determined by template matching using the ASTAR software. The predicted classes are visualized in an inverted pole Fig. (IPF). Fig. 5 shows the ground-truth labels as well as the labels predicted by the three networks each trained on one of the datasets shown in Figure 1. To better discuss the predictions, the grains in the material have been numbered in Fig. 5 a). In order to label images outside of the sample as vacuum or non-sample regions, a simple confidence threshold was used. Given the different number of classes the networks employ and the resulting differences in confidences, different thresholds had to be used in each case in order to include as many scan points of the sample as possible, while excluding most of the vacuum and non-sample regions. Comparing the predictions of the networks to the reference orientations in the IPFs, it becomes evident that all networks are generally in good agreement.

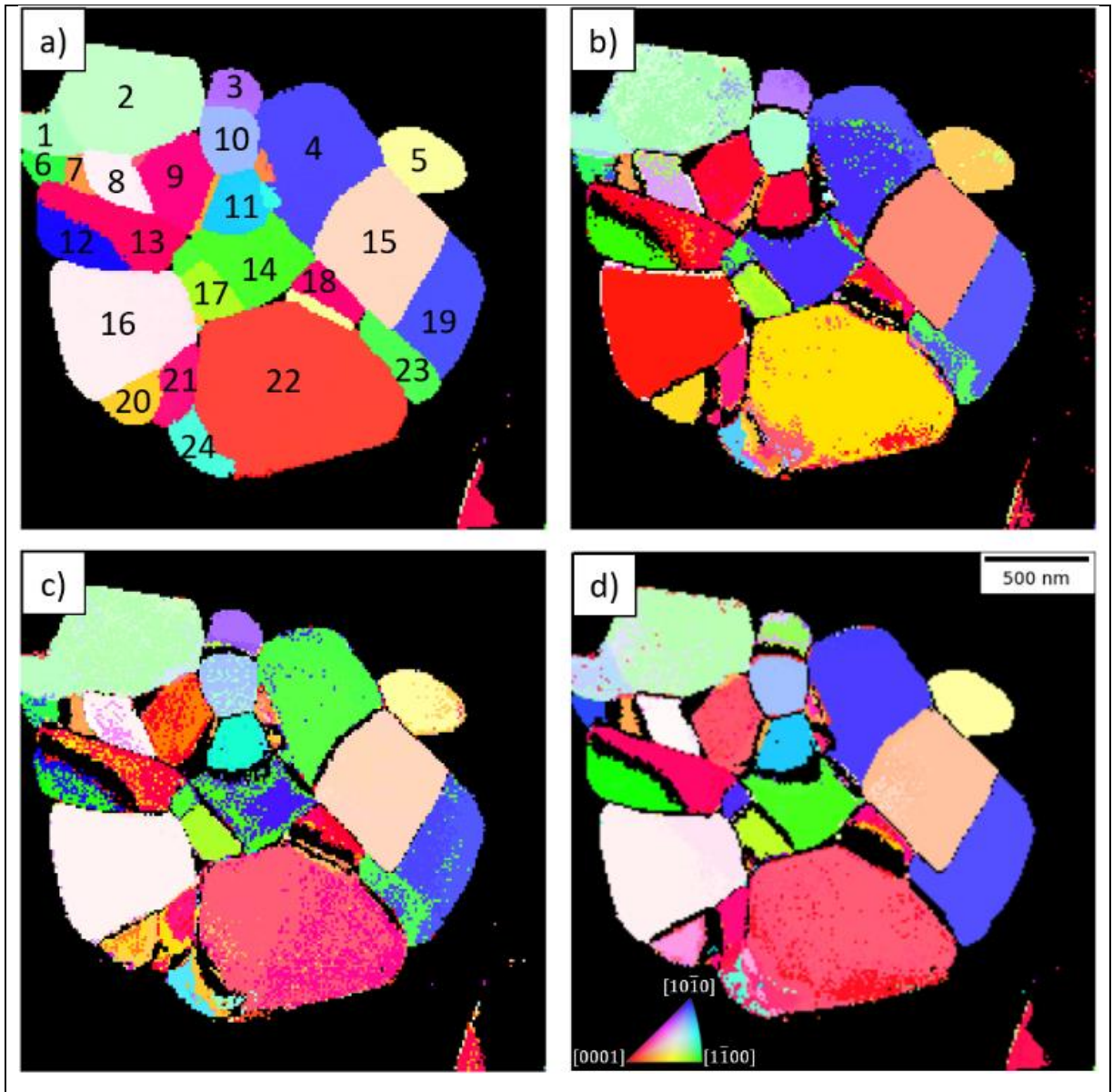

Fig. 5: Coloured IPFs of the experimental sample in the z-direction. Black areas are classified as belonging to non-sample regions. **a)** prediction of the pattern matching reference program; **b)**, **c)**, **d)** predictions of the first, second and third network trained on the datasets shown in Figure 1 respectively.

Our results can be further verified by comparing the individual diffraction pattern directly to diffraction patterns of the training datasets, that correspond to the predicted Euler angles. In Fig. 6 experimentally acquired diffraction patterns of four grains are compared to simulated non-augmented diffraction patterns of the orientations predicted by the second network (comparisons for further scan points can be found in Fig. S2 in the supplementary material). It should be noted that not all grains are equally well predicted by each network. The first network struggles the most. Here, quite a few grains are allocated significantly different orientations than in the reference. Most notably are the grains numbered 11, 16 and 22 in Fig.

5 a). This is an expected consequence of the low number of classes used for training, as it becomes likely that no class exists in the immediate vicinity of the actual orientation. Nevertheless, the accuracy of the network is high and the predictions inside of a single grain very consistent.

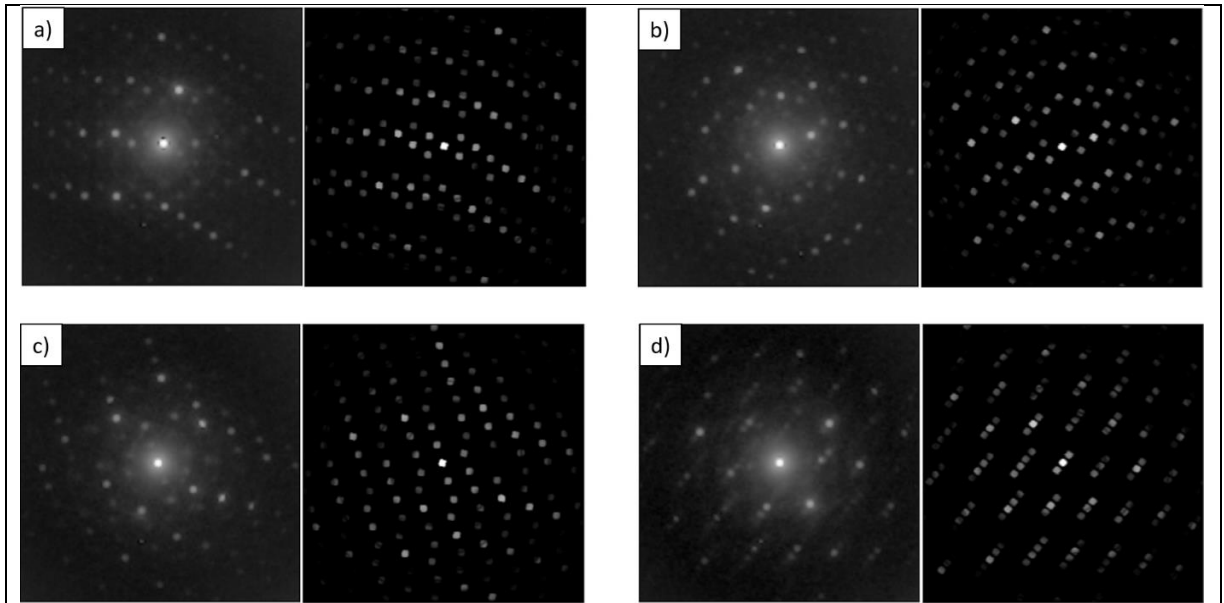

Fig. 6: Exemplary comparisons of the predictions of the network trained on the second dataset. The left pattern in each subfigure is the logarithmically scaled experimental image, while the right one is the un-augmented synthetic image, that corresponds to the predicted label. **a)**, **b)**, **c)** and **d)** correspond to scan points in grains 2, 4, 5 and 22.

The predictions of the second network show a closer resemblance to the reference orientations. However, there are more significant variations of the predictions within a single grain. This is particularly noticeable in the indecisiveness of the network when deciding between the green and blue classes for the grains numbered 4, 6, 14, 19 and 23 in Fig. 5 a). This behaviour is also visible in the predictions of the first network, though much less pronounced. In the supplementary Fig. S3, the non-augmented training images corresponding to the two classes predicted by the second network for one of the green/blue grains are overlaid. There we can see that these two predicted classes are very similar to each other, which challenges the network. When analysing the third network with the IPF images, we observe that the green/blue ambiguity as well as other intragranular uncertainties of the second network have almost entirely disappeared.

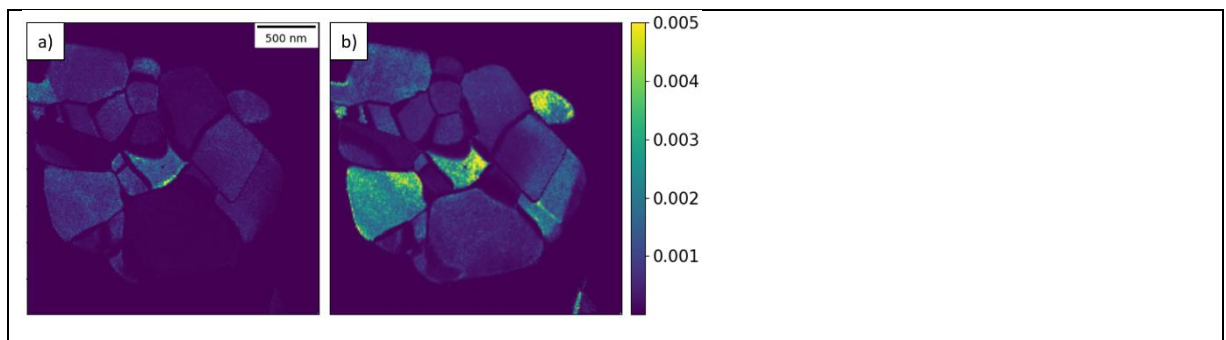

Fig. 7: Confidence of the second (**a**) and third (**b**) network in a.u., when predicting on the experimental sample. For better visualisation of the grains, the confidence was cut off at a value of 0.005.

Fig. 7 shows the confidences of the networks trained with the second and third datasets when predicting on the experimental data. The second network's confidences are a lot lower than those of the third as is expected from the significantly larger number of classes that the second network has to consider. The confidence of the third network for the right edge of the data was manually reduced, as this network assigned very high confidences to this non-sample area. Fig. S4 in the supplementary material depicts the network's confidences without the manual reduction as well as two exemplary scan points of the problematic region. Those images reveal continuous ring diffraction and strong diffuse scattering, distinctive of amorphous structures. Since our networks have not been trained on diffraction patterns of amorphous materials having no diffraction spots, hallucinatory behaviour is not surprising. In future approaches amorphous and vacuum-images should be included as separate classes for the network to predict to avoid these issues.

In order to further analyse the performance of the networks, the confidences they assign to their classes are directly visualized in IPFs. Since multiple classes are projected onto the same point in this representation, their confidences are added together. Fig. 8 shows the predictions for scan points in grains 4, 10 and 19 for all three networks. Here, the underlying source for the second network's indecisiveness is easily identified. In all three images, there are two distinct areas of high confidence. The areas are symmetric around an axis that runs horizontally through the centre of the projection. This gives credence to the idea that the issue stems from the inherent symmetry of the material system itself. This problem persists also in the predictions of the third network. However, those show a much clearer favour to one of these areas. The network has either been able to differentiate between these two symmetric solutions or internalised a bias towards one orientation. Note that the very first network does hardly exhibit any of such problems, as here only a single orientation is assigned with any amount of relevant confidence for all three scan points. This becomes even more glaring in the supplementary Fig. S5 in which the confidences of grain c) are plotted without the IPF projection.

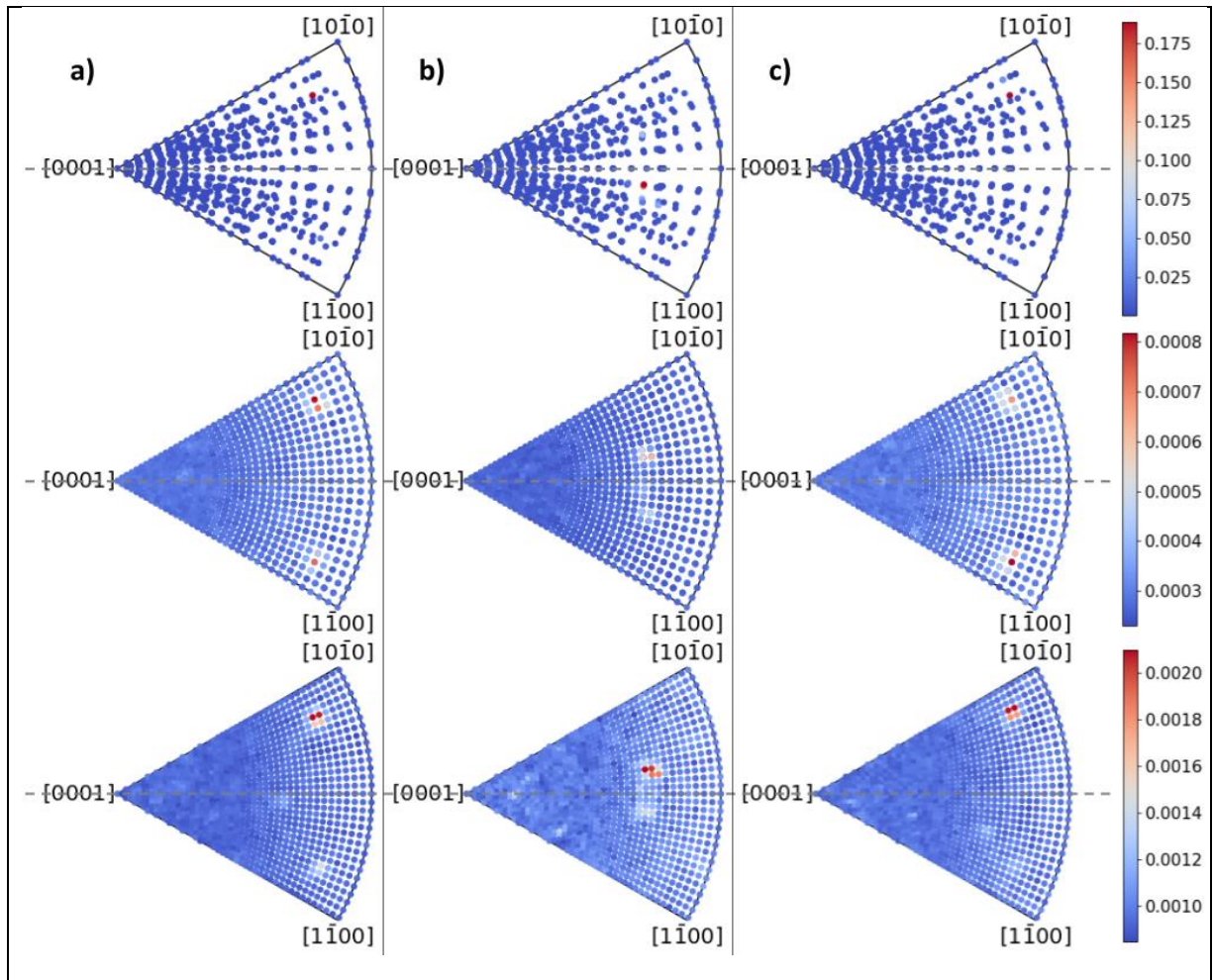

Fig. 8: Predicted probabilities of orientations for the scan points marked in Fig. 5 b). **top)** predictions of the first network. **middle)** predictions of the second network, **bottom)** predictions of the third network. **a)**, **b)**, **c)** scan points in grains 19, 10 and 4.

Diffraction pattern symmetry is a common problem when determining crystal orientations. The difference between diffraction patterns on opposite sides of the symmetry axis shown in Fig. 8 can only be simulated by incorporating dynamical diffraction. However, as such simulations are computationally expensive and the differences in the resulting intensities subtle, conventional pattern matching algorithms, like the one we used here as a reference, rely mostly on kinematic simulations for their data banks. Since our networks were trained on Bloch wave simulations that do incorporate dynamical effects, our networks might be able to learn to differentiate between otherwise similar diffraction patterns. Fig. 9 a) and b) show the proportional difference between two of those symmetrically similar orientations, i.e. $[10\bar{1}0]$ and $[1\bar{1}00]$, for both, Bloch wave simulation as well as using the a purely kinematical algorithm. For both methods, i.e. kinematic and dynamic, the diffraction spots in the first order Laue zone clearly differ for both orientations. However, in the zeroth order Laue Zone, the region of the diffraction pattern which is usually available for indexing in the experiment, there is no difference visible for the kinematic diffraction patterns (a). In contrast, in the dynamical simulation, cross-like features can be observed stemming from dynamic interferences due to double diffraction. In order to investigate the network's ability to discern the orientations through those interferences, we let the third network predict the orientations of non-augmented dynamically simulated images of the $[10\bar{1}0]$ and $[1\bar{1}00]$ orientations cropped to exclude the first order Laue zone. Fig. 9 c) shows the difference between the confidences of the network for the two images. The confidences for the vast majority classes are almost identical for both diffraction patterns and only a few orientations show a very noticeable shift. For those classes the labels have been added next to their peaks. We can see that for the majority those classes that experience a positive shift and are therefore given a higher confidence when predicting on the $[10\bar{1}0]$ image follow the pattern of φ=90°, $\boldsymbol{\varphi}$2=60°, while those classes that are given a higher probability when predicting on the $[1\bar{1}00]$ image are close to the pattern of φ=90°, $\boldsymbol{\varphi}$2=120°. As those angle pairs do correspond to the appropriate orientations, we can conclude that our third network has indeed learned to differentiate between those two directions. As to the natural question why the second network is seemingly not as competent in this regard, it may be that the high number of classes and especially the extreme density of classes close to the $[0001]$ direction preoccupy the network and therefore hinder it from focussing on learning the subtle differences between those symmetries.

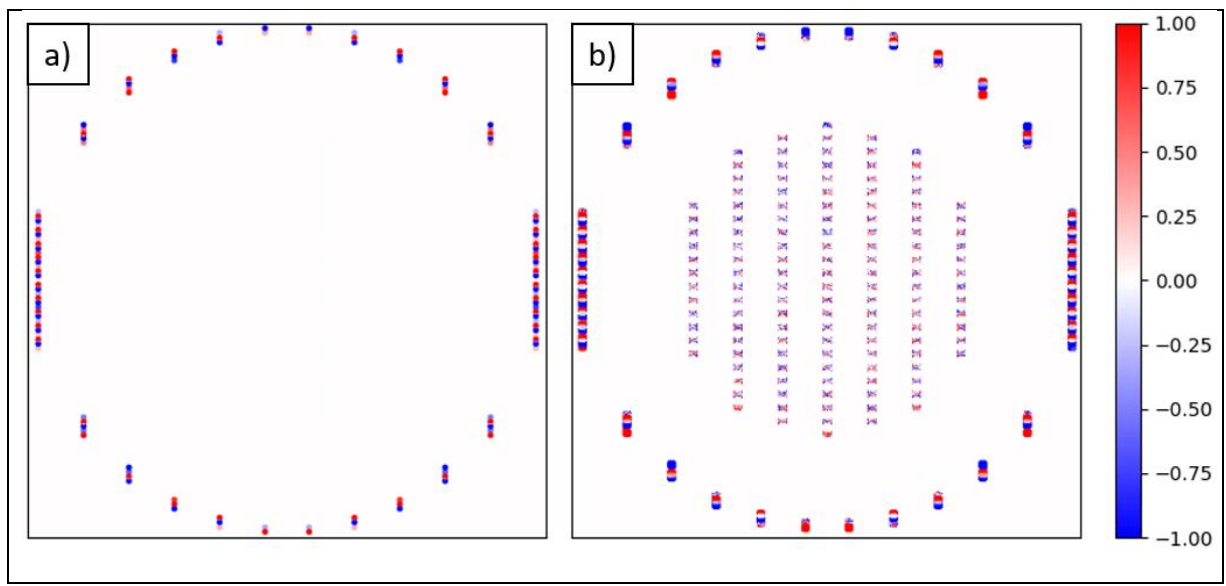

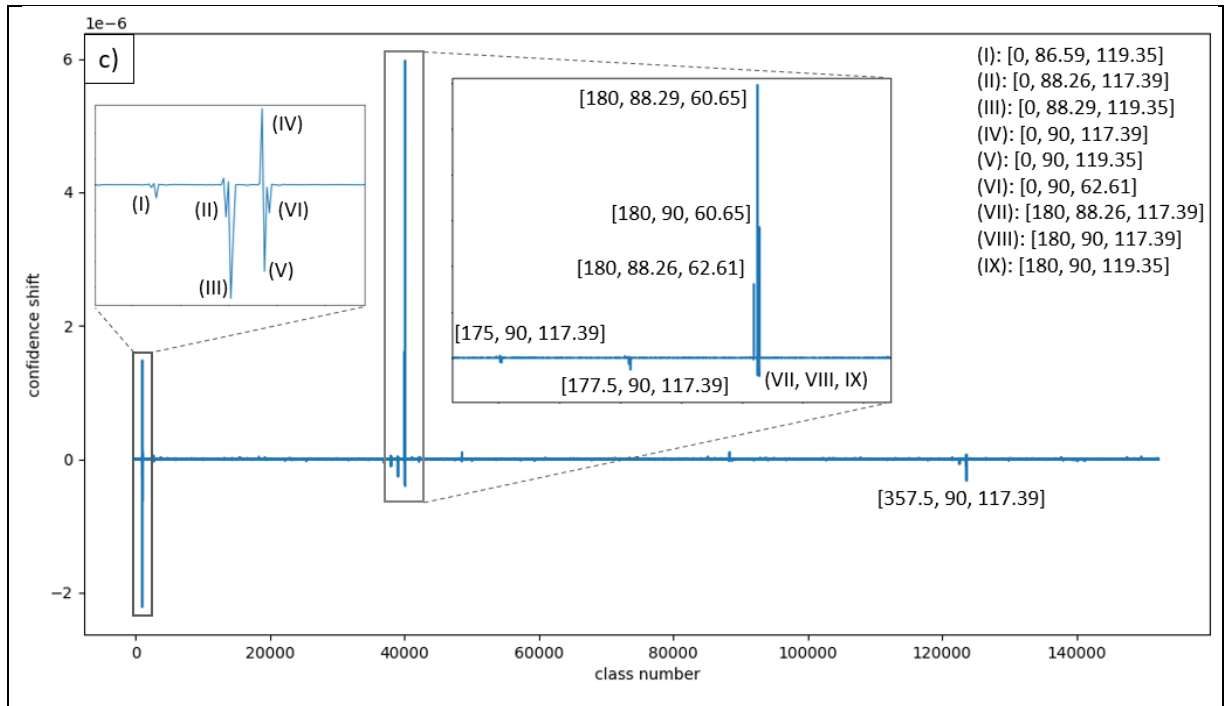

Fig. 9: Proportional change of intensity between simulated images of orientations $[10\bar{1}0]$ and $[1\bar{1}00]$ for a purely kinematic simulation **a)** and our dynamical Bloch wave simulation **b)**. **c)** difference of confidence for all classes of the third network between the prediction of non-augmented images of those two orientations that are cropped to exclude the first order Laue zone. The values in brackets are the Euler angle labels for the classes of the highest shifts.

Our analysis indicates that creating the training data in an equally-spaced manner is the most promising method for the application of neural networks, as it fills the orientation space completely without overwhelming the network with too many similar classes in certain areas. In future analysis we will therefore train solely with datasets created this way.

Additionally, we also tested our network on a non-precessed dataset of the same sample. The supplementary Fig. S6 shows a scan point of this dataset as well as the predictions of the second and third networks. This data poses an even tougher challenge, but our networks manage to give good approximation of the samples orientation.

When analysing the predictions in Fig. 5, we further notice that in some instances, the interfaces between grains are almost universally declared as non-sample regions through our threshold. Though this is in itself expected, due to the lower confidence at transitional orientation changes from one grain to another, some of these areas seem to reach significantly deep into the other grain. For further investigation a coefficient correlation map is presented in the supplementary Fig. S7 a). This was created by masking the surrounding area of the individual diffraction spots in the diffraction patterns and then comparing the resulting image to its neighbours. By doing this one can highlight the internal structure of the grains and the overlapping regions (20). Darker areas highlight diffraction patterns that are less similar to their surroundings. As expected, the grain boundaries fall into this category, however, we also see large darker areas inside some of the grains. Some of these are exemplary marked with red circles in the Fig. S7 b) showing a heat map of the number of detected diffraction spots per diffraction pattern. Here it becomes apparent that the areas in question exhibit more diffraction spots than usual. Therefore, one can conclude that these areas are the result of multiple grains overlapping or coherent grain boundaries in projection such as twin- and antiphase boundaries. Since the networks were not trained on this type of data, it is obvious that the confidence in these areas is low.

Finally, we compare the time needed to predict the orientations of all 40.000 diffraction patterns of the complete dataset. The conventional pattern matching algorithm took 108 minutes for its predictions. Our networks needed less than 2 minutes to achieve their predictions for the same dataset. This constitutes an improvement of two orders of magnitude. The high speed of the analysis of more than 300 diffraction pattern per second together with very accurate prediction results will allow in-situ applications of our method in the future, streamlining the workflow of TEM analysis considerably.

## Conclusion

In this work, we have applied convolutional neural networks to predict the crystal orientation of an $LiNiO_2$ sample by analysing the diffraction pattern obtained from a transmission electron microscope. For this, we employed classification networks with each class representing a set of three Euler angles. We created and trained with three training datasets of synthetic images with different spacing between points in angular space. By comparing the predictions with the reference data, which were acquired through pattern matching, we demonstrated that our networks achieved a high level of accuracy. This was further demonstrated by overlaying the simulated image corresponding to the predicted class directly on the experimental image. We further showed that our networks struggle with the ambiguity that arises from the symmetry of the material. This can lead to seemingly wildly varying predictions inside of a single grain of the sample. However, conventional template matching algorithms also struggle with these ambiguities and the resulting orientation maps could be subject to various orientation cleaning techniques employed in MTEX to improve the results in this regard. Fig. S8 in the supplementary material shows the result of two such techniques applied on the predictions of our second network.

Our results demonstrate that convolutional networks provide a viable way to achieve a fast and accurate orientation mapping. We also found that equal step sizes for all three Euler angles throughout the complete fundamental orientation zone provide the best predicted results. Our concept also has great promise for other material systems.

It remains to be investigated whether more complex network architectures such as transformers could increase the accuracy and certainties of the predictions especially with respect to the symmetry challenges. Furthermore, we plan to explore whether the networks could predict other parameters of the sample, such as the thickness or the amount of amorphous material as well as the sample material itself, as additional output. The current neural networks serve as an essential step towards subsequent sophisticated models to predict different phases of $LiNiO_2$ that can be formed during the recurrent electric cycling of batteries (13). These phases often maintain very similar diffraction patterns with individual spots appearing/disappearing (21) and are therefore challenging to discern for pattern matching algorithms.

In such future research, we would also test if the in-plane rotation of $\varphi 1$ should be excluded from the network predictions, as this angle could easily be determined after the other angles are found. This would decrease the number of classes and computational time significantly and free resources for finer steps in the other two angles.

## Competing interests

No competing interest is declared.

# Supplementary Information

# Determining the grain orientations of battery materials from electron diffraction patterns using convolutional neural networks

Jonas Scheunert,[1,2] Shamail Ahmed,[1,2] Thomas Demuth,[1,2] Andreas Beyer,[1,2]

Sebastian Wissel[3], Bai-Xiang Xu[3] and Kerstin Volz,[1,2]*

[1] mar.quest | Marburg Center for Quantum Materials and Sustainable Technologies, Philipps-Universität Marburg, Hans-Meerwein Straße 6, 35032 Marburg, Germany

[2]Department of Physics, Philipps-Universität Marburg, Hans-Meerwein Straße 6, 35032 Marburg, Germany

[3]Mechanics of functional materials, TU Darmstadt, Otto-Berndt Straße 3, 64287 Darmstadt, Germany

*Corresponding author: kerstin.volz@physik.uni-marburg.de

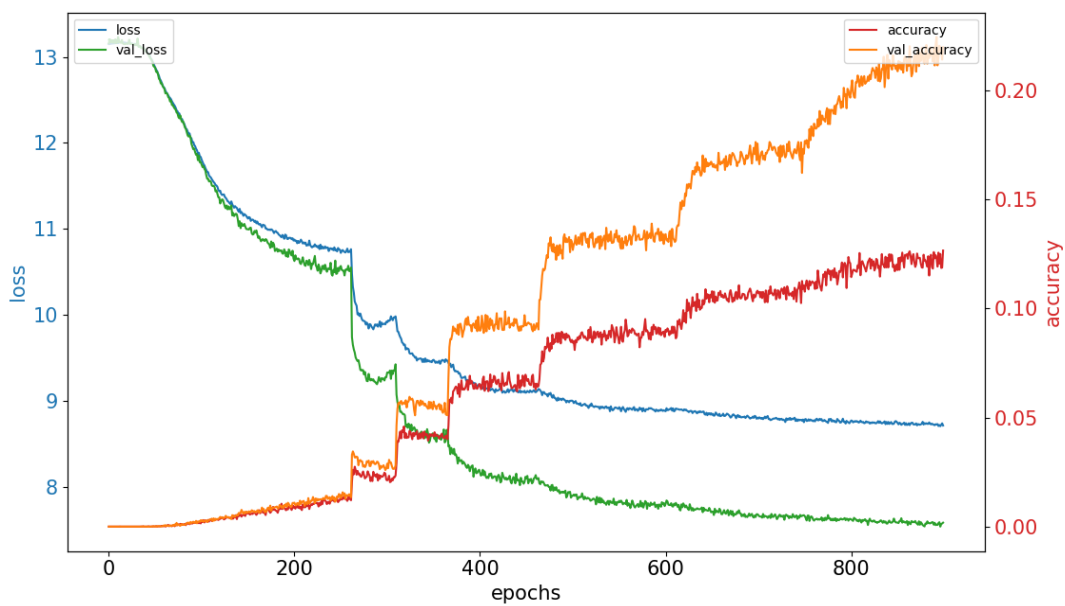

S1: Accuracy and loss tracked for both the training and validation dataset during the training process of the second network. The sudden changes in both metrics correspond to an adaptation of the learning rate after the validation loss was not reduced for 20 epochs. The loss and accuracy being both higher and lower respectively than their validation counterparts is the result of our dropout layers that are only active during the training process.

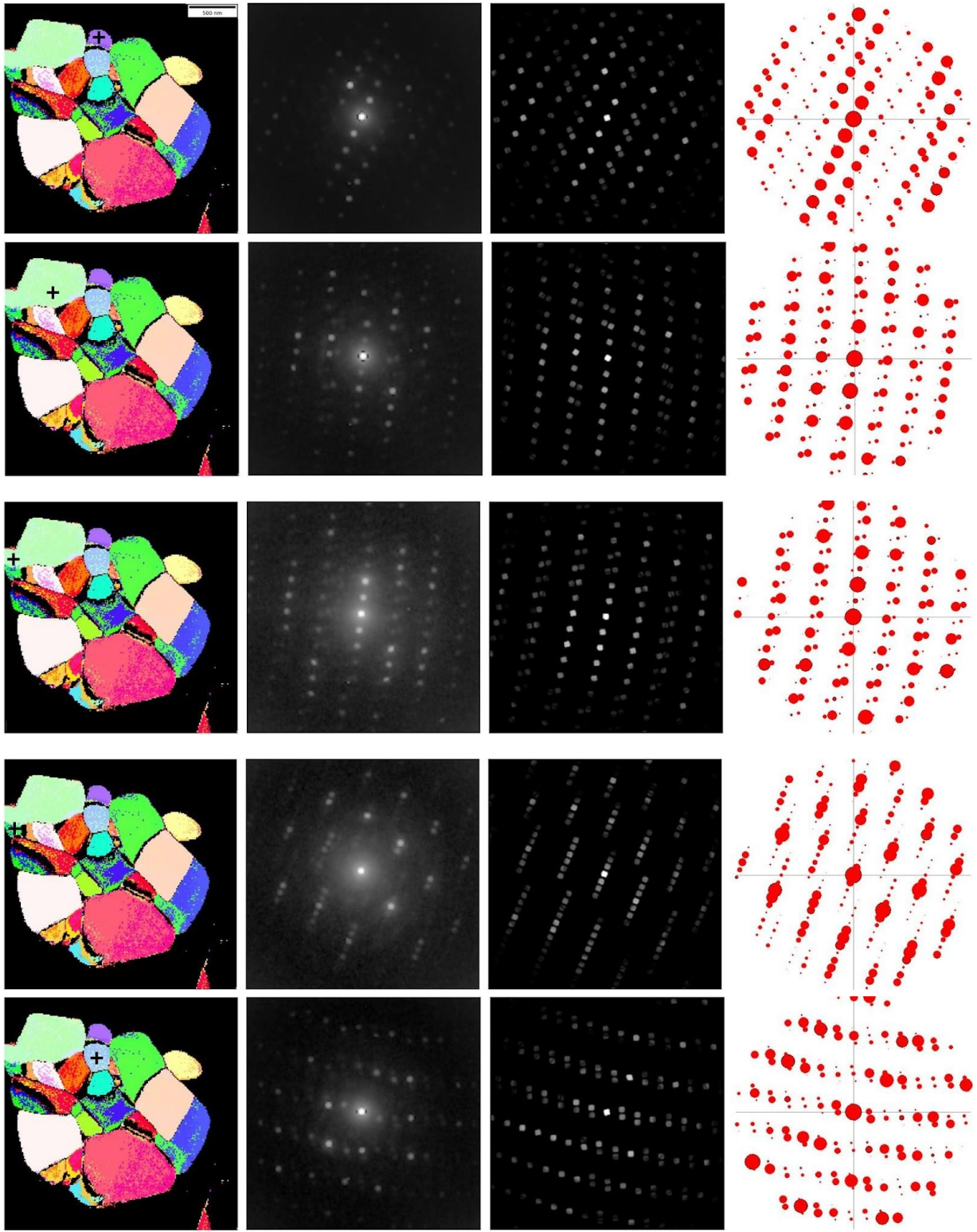

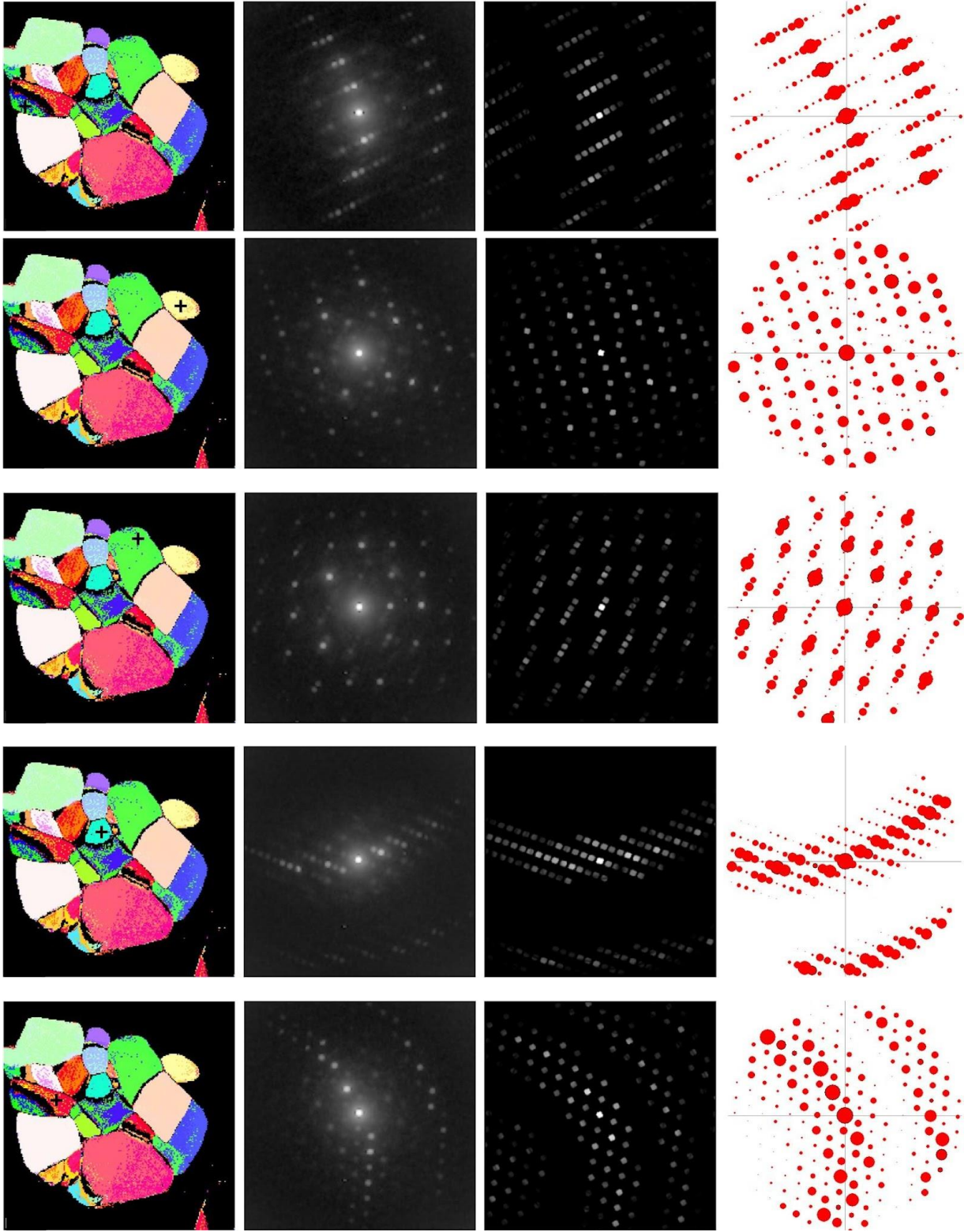

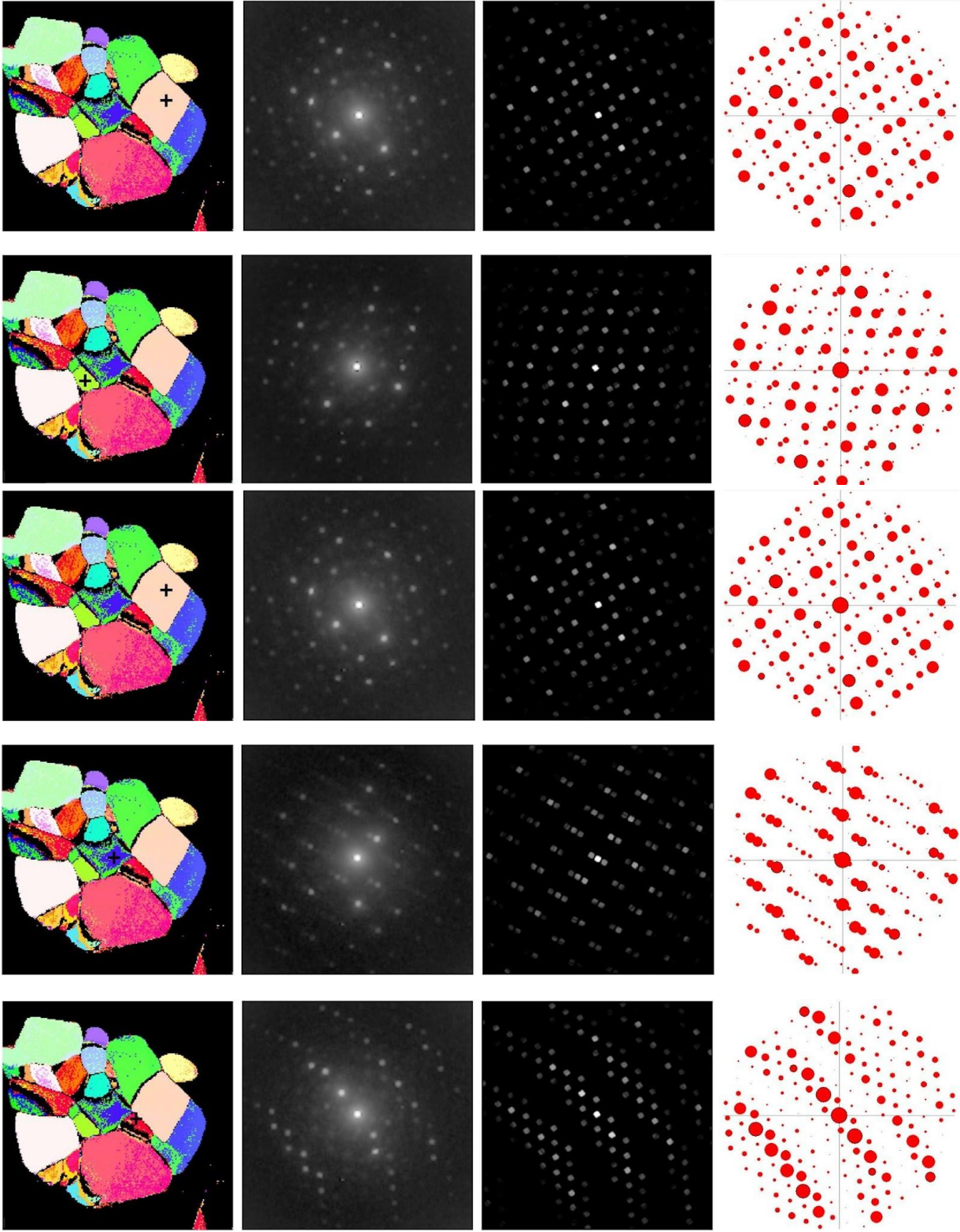

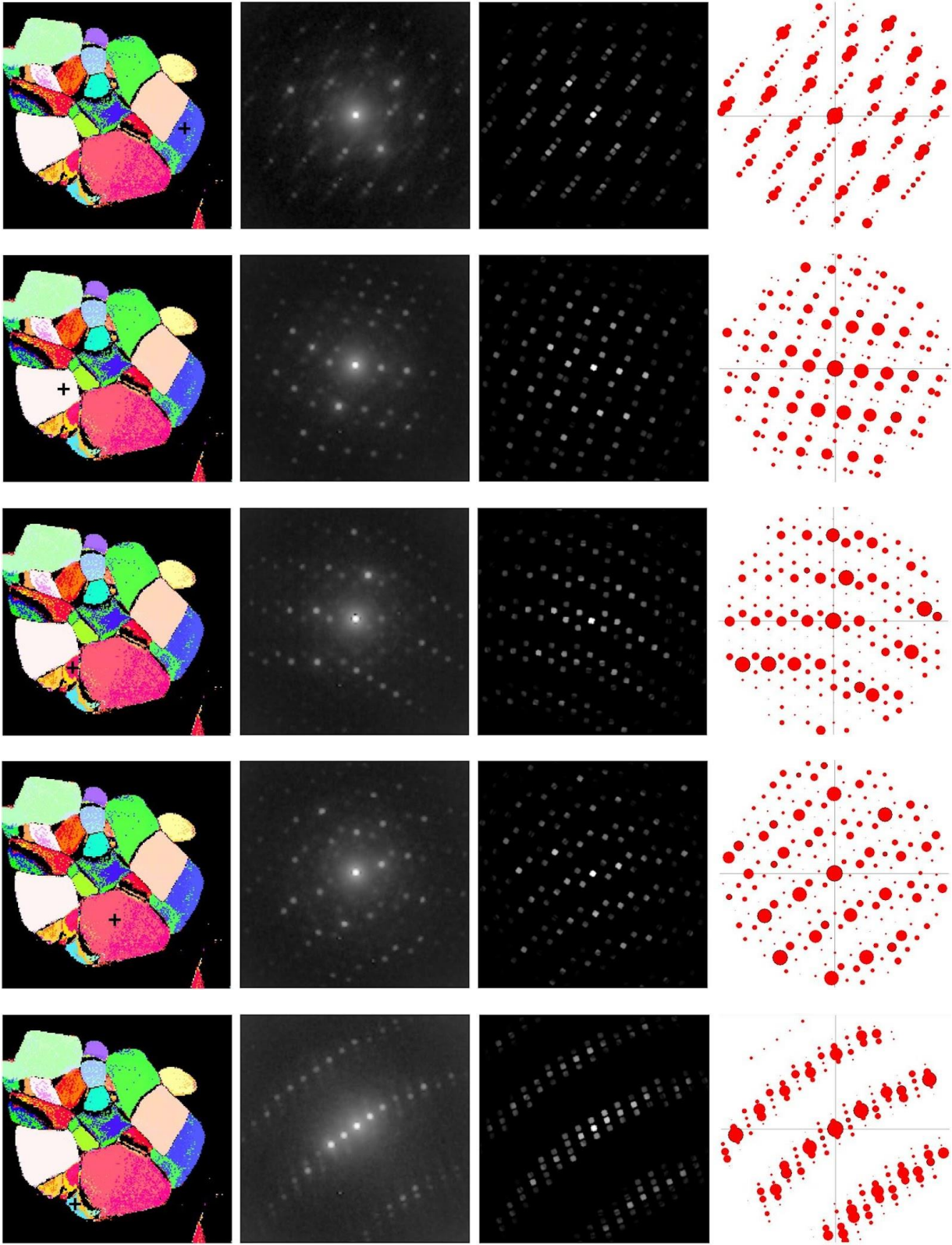

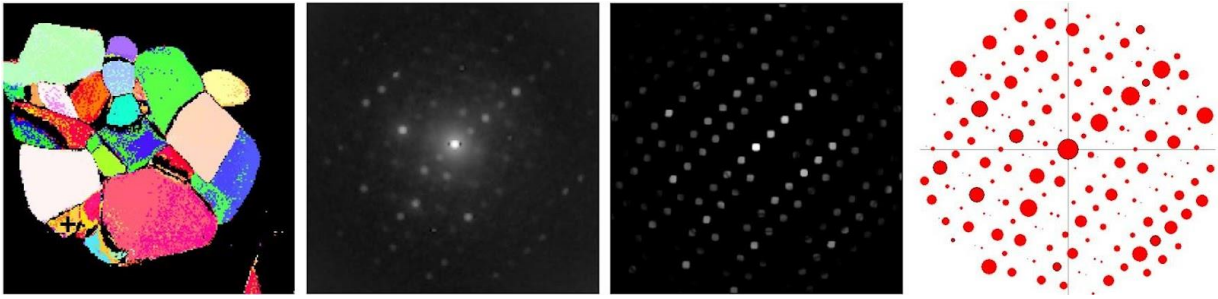

S2: Comparison between various STEM-images and the simulated images that correspond to the predictions. From left to right: IPF with a cross marking the scan point; STEM-image; BW simulated image corresponding to the prediction of the second network, kinematic simulation corresponding to the prediction of the ASTAR program.

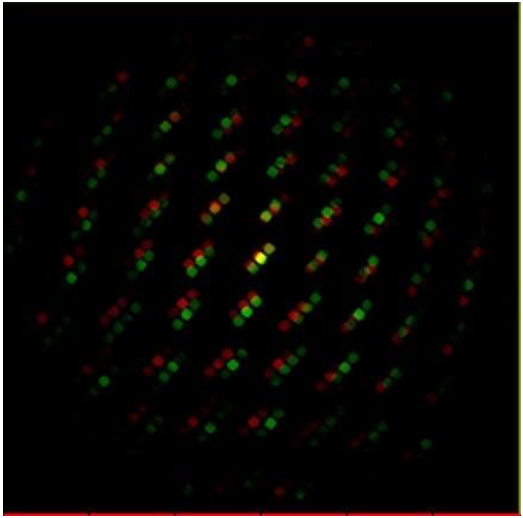

S3: Overlay of two simulated non-augmented diffraction patterns corresponding to the two predicted classes of a green/blue grain by the second network.

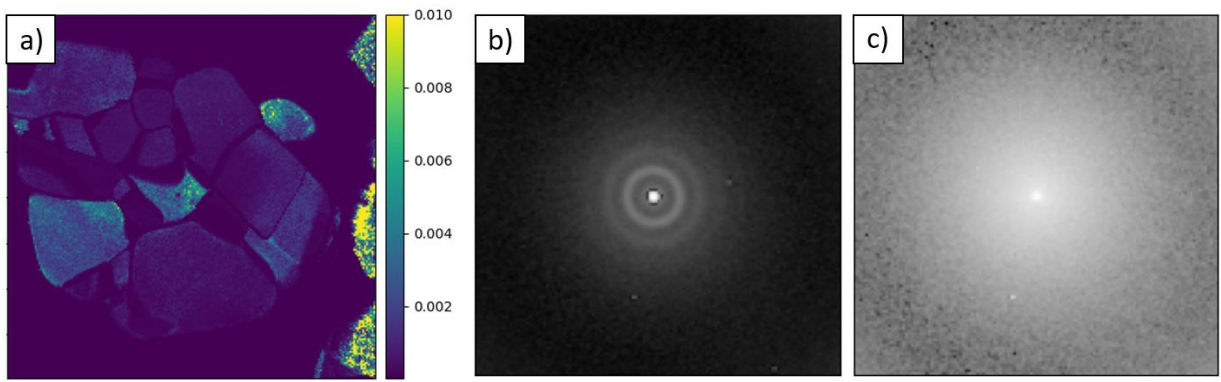

S4: **a):** confidences of the third network without suppressing the right edge of the data; **b)** and **c)** scan points inside the problematic region.

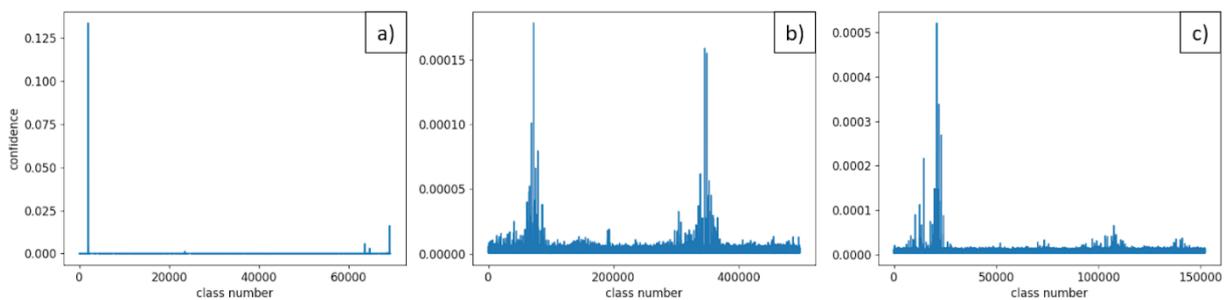

S5: Predicted probabilities of all classes for scan point a, marked in Fig. 6 b). **a), b)** and **c)** correspond to the first, second and third network.

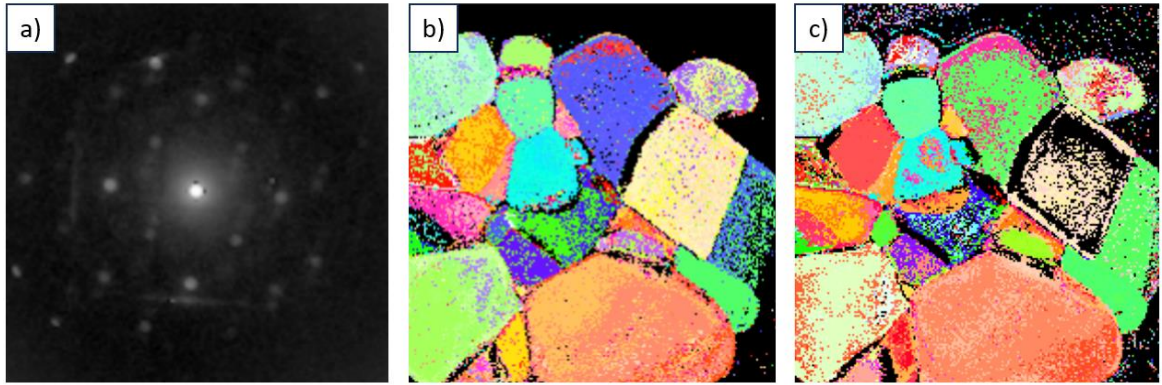

S6: **a)** non-precessed diffraction pattern. **b), c):** predictions of the second and third networks of the non-precessed dataset.

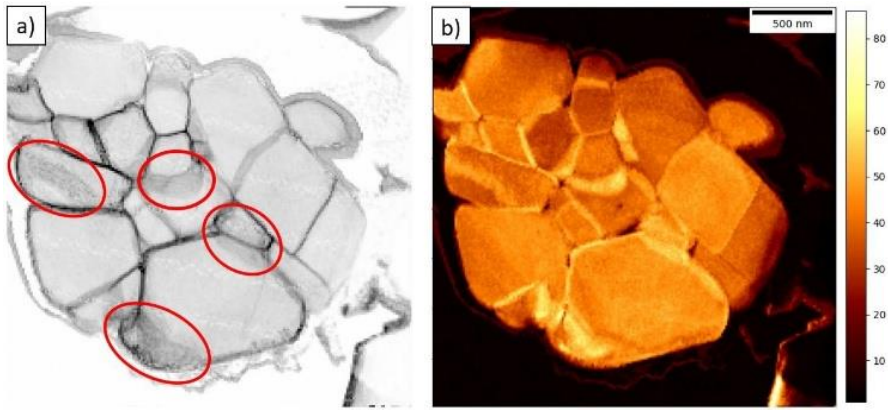

S7: **a):** Coefficient correlation map, red circles indicate exemplary large areas of significant grain overlap; **b):** heat map of the number of detected diffraction spots.

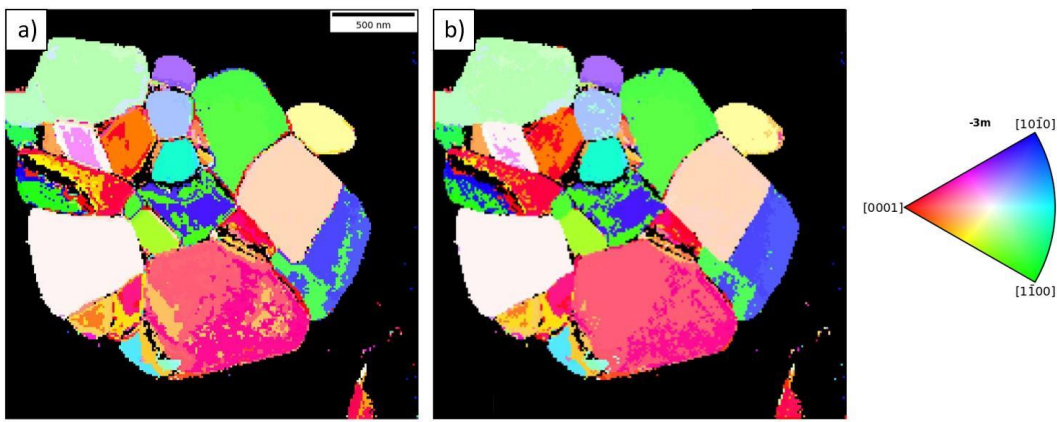

S8: IPF visualization of the predictions of the second network after post processing. **a):** The individual diffraction patterns were averaged with their eight surrounding scan points to produce a cleaner diffraction image. **b):** The confidences for each class for the surrounding scan points were weighted with 0.125 and then added to the confidences of the central scan point to create more uniform predictions for individual grains.